\documentclass[12pt]{article}
\usepackage{epsfig,graphicx, amssymb}
\evensidemargin  \oddsidemargin
\usepackage{amsmath}

\begin{document}

{\large
\begin{center}

PULSIVE FEEDBACK CONTROL FOR STABILIZING UNSTABLE
PERIODIC ORBITS IN A NONLINEAR OSCILLATOR WITH A NON-SYMMETRIC POTENTIAL
\\ ~\\ ~\\
G. LITAK$^1$, M. ALI$^2$, L.M. SAHA$^3$
\\~ \\
$^1$Department of  Applied Mechanics, Technical University
of
Lublin,   
Nadbystrzycka 36, PL-20-618 Lublin, Poland \\~\\
$^2$Department of Mathematics, Faculty of Mathematical
Science, University
of
Delhi, Delhi 110007, India \\~\\
$^3$Zakhir Husain College, University
of
Delhi,  Delhi 110002, India
\end{center}

\vspace{2cm}

\noindent {\bf Abstract} \\
We examine a strange chaotic attractor and its unstable periodic orbits in 
case of 
one degree of freedom nonlinear oscillator with non symmetric potential. 
We propose an efficient method of chaos control stabilizing these orbits 
by a pulsive feedback technique. Discrete set of pulses enable us to 
transfer the system from one periodic state to another. }

\newpage

A chaotic motion appearing in specific physical systems may show various 
positive and negative effects [Thomsen 2003] depending on 
applications. 
Opportunity of its 
control triggered a new filed of nonlinear research [Fradkov \& 
Evans 2005]. 
Starting with a strange chaotic attractor to system control is a comfort 
situation because of infinite number of unstable periodic orbits
included in it.  
This novel idea of using unstable periodic orbits to chaos control has 
been invented by  Ott, Geborgi and Yorke [1990] (OGY method)  
and applied to many physical systems [Fradkov \& Evans 2005]. 
Namely, Ott {\em et al.} [1990] have shown that one can convert a 
chaotic
 attractor  to any of a large number of possible attracting time-periodic
 motions by making only small time-dependent perturbations of  
available
 system parameters. In other words one can stabilize any unstable periodic
 orbit included in a strange chaotic attractor.
The next important step was the self-controlling feedback method
 introduced by Pyragas [1992], where the small perturbations 
were 
continuous
 in time. 

Impulsive methods for dynamical systems' control and synchronization are
 some known approaches in the field of chaos [Yang {\em et al.} 1997,
Osipov {\em et al.} 1998a, Osipov {\em et al.} 1998b, Sun \& Zhang 2004,
Sun {\em et al.} 2004, Khadra {\em et al.} 2005]. It 
was used
 successfully for controlling R\"{o}ssler system [Yang {\em et al.} 1997] 
and the Duffing
 oscillator [Osipov {\em et al.} 1998a] to periodic motions. More recent 
paper  about impulsive
 control was more successful in establishing more conservative and
 sufficient conditions for the stabilization and synchronization of Lorenz
 systems via impulsive control. In their recent work, Sun \& Zhang [2004], 
 presented some new theorems on the stability of impulsive control
 systems, which was applied successfully to the Chua's oscillator. Based
 on stability theory of impulsive differential equation and new comparison
 theory, the authors of [Sun {\em et al.} 2004] studied the chaos 
impulsive synchronization of
 two coupled chaotic systems using the unidirectional linear error
 feedback scheme. Moreover, in the most recent work [Khadra {\em et al.} 
2005], this approach was
 used with non-linear partial differential equations. The authors
 determined a criterion for the solutions of these partial differential
 equations to be equi-attractive in the large and estimated the basin of
 attraction in terms of the impulse durations and the magnitude of the
 impulses. In our paper, we apply the same impulsive method with a linear
 feedback strategy based on the knowledge of unstable periodic orbit
embedded within the chaotic attractor of the original system.

We start from a single degree of freedom 
system 
subjected 
to an external excitation
with a non-symmetric
stiffness given by the following   equation:
\begin{equation}
\label{eq1}
\ddot{x} + \alpha  \dot{x}
+ \delta x +\gamma x^2=\mu \cos{ \omega t}
\end{equation}
where $x$ is a displacement,  $\alpha \dot{x}$ is linear damping,
$\mu \cos{\omega t}$ is an external excitation, while $\delta x$ and 
$\gamma x^2$
are  linear and  quadratic force terms.

The above equation has been extensively studied by 
Thompson \& Hunt [1989], who found chaotic  behavior there and 
examined 
transitions to chaos  through a 
a global homoclinic 
bifurcation and a cascade of period doubling bifurcations just before
escape from the potential well. 
Such systems (Eq. \ref{eq2}) have been also a subject 
of studies for many other researchers, inspired by  possible 
applications in description of mostly mechanical systems 
[Szabelski \& Samodulski 
1985, 
Szemplinska-Stupnicka \& 
Rudowski 1993, Szemplinska-Stupnicka 1995, Rega {\em 
et al.} 1995, Rand 2003, Litak  {\em et al.} 2005, Litak  {\em et al.} 
2006] 
and the catastrophe 
theory 
[Poston 1978]. 
They were also linked to possible meta-stable states of atoms and 
they appeared in problems within the elastic theory 
[Thompson 1989,Thompson \& Hunt 1984].

The above equation is equivalent to the following autonomous system of 
three
first-order differential equations:
\begin{eqnarray}
\dot{x}&=&y \nonumber \\
\dot{y} &=& -\alpha y - \delta x - \gamma x^2 + \mu \cos ( \omega z) 
\label{eq2} \\
\dot{z}&=& 1, \nonumber
\end{eqnarray}
where $z=t$, therefore, whenever one attempts to integrate the system (Eq. 
\ref{eq2}), 
one must pay attention to the fact that the initial conditions must be 
such 
that
$z_0=t_0$.
 From the recent paper by Litak {\em et al.} [2006] it is clear 
that 
the system given by Eq. \ref{eq1}
 exhibits a chaotic motion and a strange attractor for the 
parameters' values $\alpha=1.0$, $\delta=1.0$, $\omega=0.85$, 
$\gamma=1.089$ and $\mu=0.608$.
For practical reasons of system control in relatively large values of 
variables we have rescaled the variables by 
acquiring the following variables' changes:

\begin{eqnarray}
x \rightarrow \frac{x}{100}, \nonumber \\
y \rightarrow \frac{y}{100}. \label{eq3}
\end{eqnarray}
After the above transformation the system (Eq. \ref{eq2})  remains 
the 
same while the parameters'
values of $\gamma$ and $\mu$  would become 0.01089 and 60.8, respectively.
 The strange attractor of a chaotic motion of this system has been  shown 
in Fig. \ref{fig1}.
We can start examining  the system Eq. \ref{eq2} looking for an 
unstable 
periodic orbit: 
\begin{equation}
X^*(t)=[x^*(t),y^*(t),z^*(t)] \label{eq4}
\end{equation} 
embedded 
within its
 chaotic attractor of period $2 \pi / \omega$. This orbit has been 
obtained 
numerically by a
method of recurrence and has been shown in Fig. \ref{fig2}.
The basic idea of recurrence is to
 wait two successive iterations of the designed Poincare map of sections 
to fall in a sufficiently small neighborhood.
 In our case, for the sake of more accuracy, we have used the same concept 
but with a
little modification.
 Given the dimension of the phase space and the range of the variables, we 
where able to
 determine a rectangle in the phase space where points of the unstable 
periodic orbit are
suspected to be within it.
 Using Mathematica, we were able to develop a code that can detect a 
smaller rectangle
 within the previous one in which points of the unstable periodic orbit 
are lying within
 it. This was done by dividing the previous rectangle over a net of 10000 
smaller and
 identical rectangles then integrating the given flow starting with each 
mesh on the net
 and finding the mesh at which the smallest recurrence occurs.
 Repeating the same procedure successively finitely many times, we were 
able 
to determine a
 point (mesh) at which an arbitrarily small recurrence occurs.
 Integrating and plotting the orbit initialized at this mesh over the same 
period of the
 Poincare map, would give us the best approximation of the required 
unstable periodic
orbit of the given period.

 Using a feedback technique, we have been able to stabilize the
 unstable period-1 orbit of this system which was embedded
 within its chaotic attractor (Fig. \ref{fig2}). This has been  done
by adding a small perturbation: 
\begin{equation}
\epsilon(X(t)-X^*(t))
\end{equation}
to the considered system (Eq. \ref{eq2}).
The stabilized period-1 orbit of the given system is shown in Fig. 
\ref{fig3}. Starting with the same initial conditions as in unperturbed 
system (Fig. \ref{fig1}) we have easily obtained periodic motion of the 
system.  
In 
this case control parameter was assumed to be $\epsilon=-0.5$. For smaller 
$|\epsilon|$  the same final aim have been  obtained but in relatively
longer time. In fact  the resulting behavior of the system does not 
change qualitatively while changing a value of $\epsilon$.

 An alternative method  is to apply 
the same
 feedback technique to the given system but on a discrete scale as pulses
 (pulsive feedback technique). 
For instance, one does not need to keep watching and observe the dynamics
 of the system all the time in order to supply the system with necessary
 perturbations continuously, instead this task is done at predetermined
 discrete and equal time intervals proportional to the period of the
 unstable periodic orbit that we wish to stabilize. In the example given
 below, each pulse is set to last for a time equal to the
 pre-assigned integration step and several pulses are provided to the 
system
 per period. This would eliminate all waste of unnecessary energy
 required in the original continuous feedback technique and make it an
 optimal technique as the cost of control is minimized.
For the integration, we used a Runge-Kutte method of second order with a 
 step size equal to $2\pi/(1000 \omega)$ i.e. one cycle of period $\tau=2 
\pi /\omega$  was 
divided into  1000 equal
time intervals for the integration purpose.

Applying the pulsive feedback technique to the examined system (Eqs 2-3) 
in case of the time interval 
between two
 successive pulses was  $\pi/(3\omega)$ produced interesting results. 

Beside the fact that we have been 
able to
 stabilize the same unstable period-1 orbit of this system, we have   
also been able
 to transfer the system from one periodic state to another by varying the
 pulses strength $\epsilon$. Figures 4a-c   show stabilized 
period-1 ($2 \pi/\omega$),
 period-2 and period-4 orbits of the system (Eq. \ref{eq2}) using the 
pulsive 
feedback
 technique with $\epsilon=-0.05$ , $\epsilon=-0.03$  and 
$\epsilon=-0.009$, respectively.
 In fact, we have found that the system (Eq. \ref{eq2}) with pulsive 
feedback undergoes a
 period doubling bifurcation as the pulses strength $\epsilon$ increases 
in the
 interval $[-0.05,0]$.
A bifurcation diagram of our system  with pulsive feedback control
is shown in Fig. \ref{fig4}c with $\epsilon \in [-0.05,0]$.  
This is the principal result of our present investigations. 
We would like to stress that it is very promising for 
engineering practice and 
real life 
experiments
 as it enables switching between one state and another of the same
 system by simple controlling the pulses' strength provided for the
 system. 

Here using the same predetermined unstable periodic orbit of period  
embedded
 within the chaotic attractor, we have  not only been able to 
stabilize this   
 orbit but also a variety of other unstable periodic orbits, as shown in 
 the bifurcation diagram (Fig4d). The period and nature of the stabilized 
orbit  
 depend entirely on the perturbation's coefficient  (which is supposed to 
 be small). Different stabilized orbits are shown in Figs 4a-c In 
 order to stabilize an unstable periodic orbit of a dynamical system,   
 knowledge of the same orbit is not necessary, instead, one may make use
 of any other unstable periodic orbit of lower period available of the
 same system.

The above result may be also considered as an advantage of the pulsive 
feedback
 technique over the well known feedback technique 
[Pyragas 1992, Fradkov \& Evans 2005] which does not enable
 such switching from one state a dynamical system to another without
 re-engineering the whole method in order to suit a particular state.
In this paper we have applied this technique to a system with single 
non-symmetric well [Thompson \& Hunt 1989]
potential. However we claim that 
it should work successfully for any system with a period doubling 
bifurcation cascade. Note also, our approach to system control differs 
from so called impulsive control methods 
[Yang {\em et al.} 1997, Osipov {\em et al.} {1998a, Osipov {\em et 
al.} 1998b] where
the main issue was suppressing chaotic motion. In their cases 
the system is stabilized by adequately strong impulsive signal  which
drive it to periodic motion of various properties. These periodic orbits, 
in contrast to the present consideration, 
were not related to unstable periodic orbits of a strange attractor. In 
our case the pulsive control method is based on unstable periodic orbit 
and it is making use of a feedback control in discrete way and 
the stabilized orbits are strongly related to other unstable orbits of a 
strange attractor with multiple period. 

\section*{Acknowledgments}
GL would like to acknowledge a partial support from the Polish State 
Committee of Scientific 
Research.

\newpage

\section*{References}
~

Fradkov, A.L. \& Evans, R.L. [2005] "Control of chaos: Methods
and applications in engineering", {\em Annual Reviews in Control} 29,
33--56.

Khadra, A. Liu X. \& Shen, X. [2005] "Impulsive control and
synchronization of spatiotemporal
chaos", {\em Chaos, Solitons \& Fractals}, 26, 615-636.

Litak, G., Syta, A., Borowiec, M. \& Szabelski, K. [2005]   
"Transition to chaos  in the self-excited system with
a cubic double well potential and parametric forcing", {\em Int. J. 
Non-Linear
Mech.}, submitted.

Litak G, Syta A \& Borowiec M. [2006] "Suppression of chaos by 
weak resonant excitations in a nonlinear oscillator with a non-symmetric
potential", {\em Chaos, Solitons  \& Fractals}, in press.

Osipov, G., Glatz, L. \& Troger, H. [1998a]  "Suppressing chaos in the
Duffing oscillator by impulsive actions", {\em Chaos Solitons \& Fractals}
9, 307--321.

Osipov, G.V., Kozlov, A.K. \& Shalfeev V.D. [1998b] "Impulsive control of 
chaos in continuous systems", {\em Phys. Lett.} A 247, 119--128.

Ott, E., Geborgi, C. \& Yorke, J.A. [1990] "Controlling chaos", {\em Phys.
Rev. Lett.} 64
1196--1199.

Poston, T. \& Stewart J. [1978]  {\em Catastrophe theory and its
applications}, (Pitman, London)

Pyragas, K. [1992] "Continuous control of chaos by
self-controlling feedback",
{\em Phys. Lett.} A 170, 421--428.

Rand, R.H. [2003] {\em Lecture notes on nonlinear vibrations}
(Ithaca:The Internet-First University Press, Ithaca), 
http://www.tam.cornell.edu/randdocs/.

Rega, G., Salvatori, A. \& Benedettini, F. [1995] "Numerical and
geometrical analysis of bifurcation and chaos for an asymmetric elastic
nonlinear oscillator", {\em Nonlinear Mechanics} 7, 249--272.

Sun, J.T, Zhang, Y.P. \& Wu, Q.D. [2002] "Impulsive Control for the
Stabilization and Synchronization of
Lorenz Systems". {\em Phys. Lett.} A, 298, 153--160.

Sun, J.T \& Zhang, Y.P. [2004] "Impulsive Control and Synchronization of
Chua's Oscillators".
{\em Mathematics and Computers in Simulation}, 66 499--508.

Sun, J.T., Zhang, Y.P., Qiao F. \& Wu, Q.D. [2004] "Some Impulsive
Synchronization Criterions for Coupled
Chaotic Systems via Unidirectional Linear error Feedback Approach",
{\em Chaos,
Solitons \& Fractals}, 19, 1049--1055.

Szabelski, K. \& Samodulski, W. [1989] "Drgania ukladu z
niesymetryczna charakterystyka sztywnosci przy parametrycznym i
zewnetrznym wymuszeniu", {\em Mechanika Teoretyczna i Stosowana} 
23, 223--238.

Szempli\'nska-Stupnicka, W. \& Rudowski, J. [1993]
"Bifurcattions phenomena in a nonlinear oscillator: approximate analytical
studies versus computer simulation results", {\em Physica} D 
66, 368--380.

Szempli\'nska-Stupnicka W. [1995] "The analytical
predictive criteria for chaos and escape in nonlinear oscillators: A
survey", {\em Nonlinear Dynamics} 7 129--147.

Thompsen, J.J. [2003] {\em Vibrations and stability} (Springer, Berlin). 

Thompson, J.M.T. \& Hunt, G.W. [1984] {\em Elastic instability
phenomena},
(Wiley, Chichester).

Thompson, J.M.T. [1989] "Chaotic Phenomena Triggering the
Escape from a Potential Well", {\em Proceedings of the Royal Society of 
London} A 421, 195--225.

Yang, T., Yang, C-M. \& Yang, L-B. [1997] "Control of R\"{o}ssler system 
to periodic motions using impulsive control methods",  
{\em Phys. Lett.} A 232,
356--361.

\newpage

\begin{figure}[htb]
\hspace{0.5cm}
\begin{center}
\epsfig{file=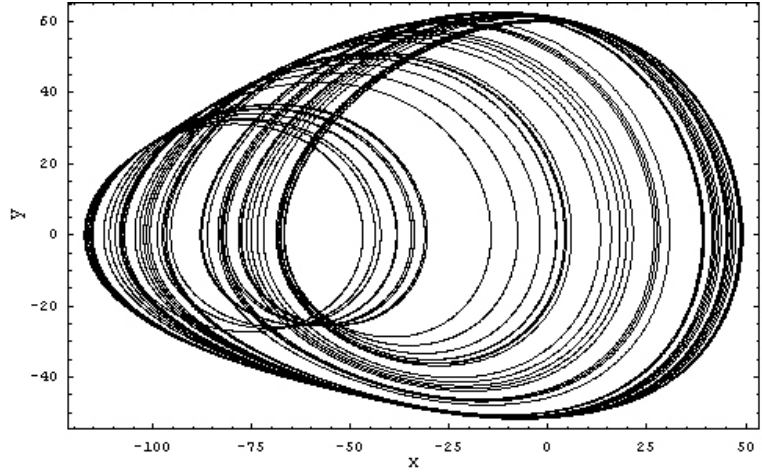,width=10.2cm,angle=0}
\end{center}
\caption{ \label{fig1}
The chaotic strange attractor of the system (Eq. \ref{eq2})  for the
parameters
values $\alpha=1.0$, $\delta=1.0$, $\omega=0.85$,
$\gamma=0.01089$ and $\mu=60.8$.}
\end{figure}

\begin{figure}[htb]
\hspace{0.5cm}
\begin{center}
\epsfig{file=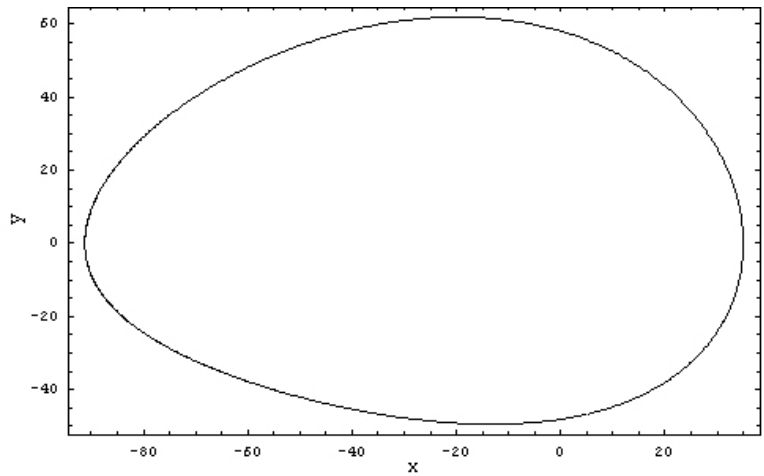,width=10.2cm,angle=0}
\end{center}
 \caption{ \label{fig2}
An unstable period-1 orbit of the strange attractor Fig. \ref{fig1}.}
\end{figure}

\begin{figure}[htb]
\hspace{0.5cm}
\begin{center}
\epsfig{file=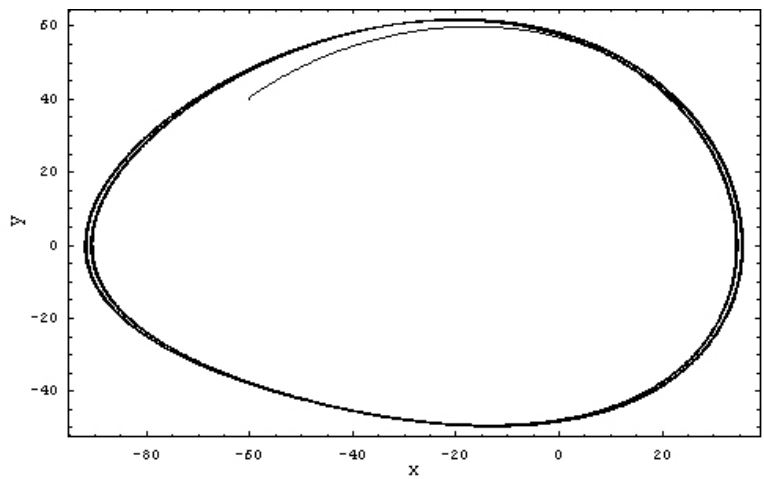,width=10.2cm,angle=0}
\end{center}
 \caption{ \label{fig3}
Stabilized period-1 orbit of system (Eqs. \ref{eq2})
 using the feedback technique with $\epsilon=-0.5$.}
\end{figure}

\begin{figure}[htb]
\hspace{0.5cm}
\begin{center}
\epsfig{file=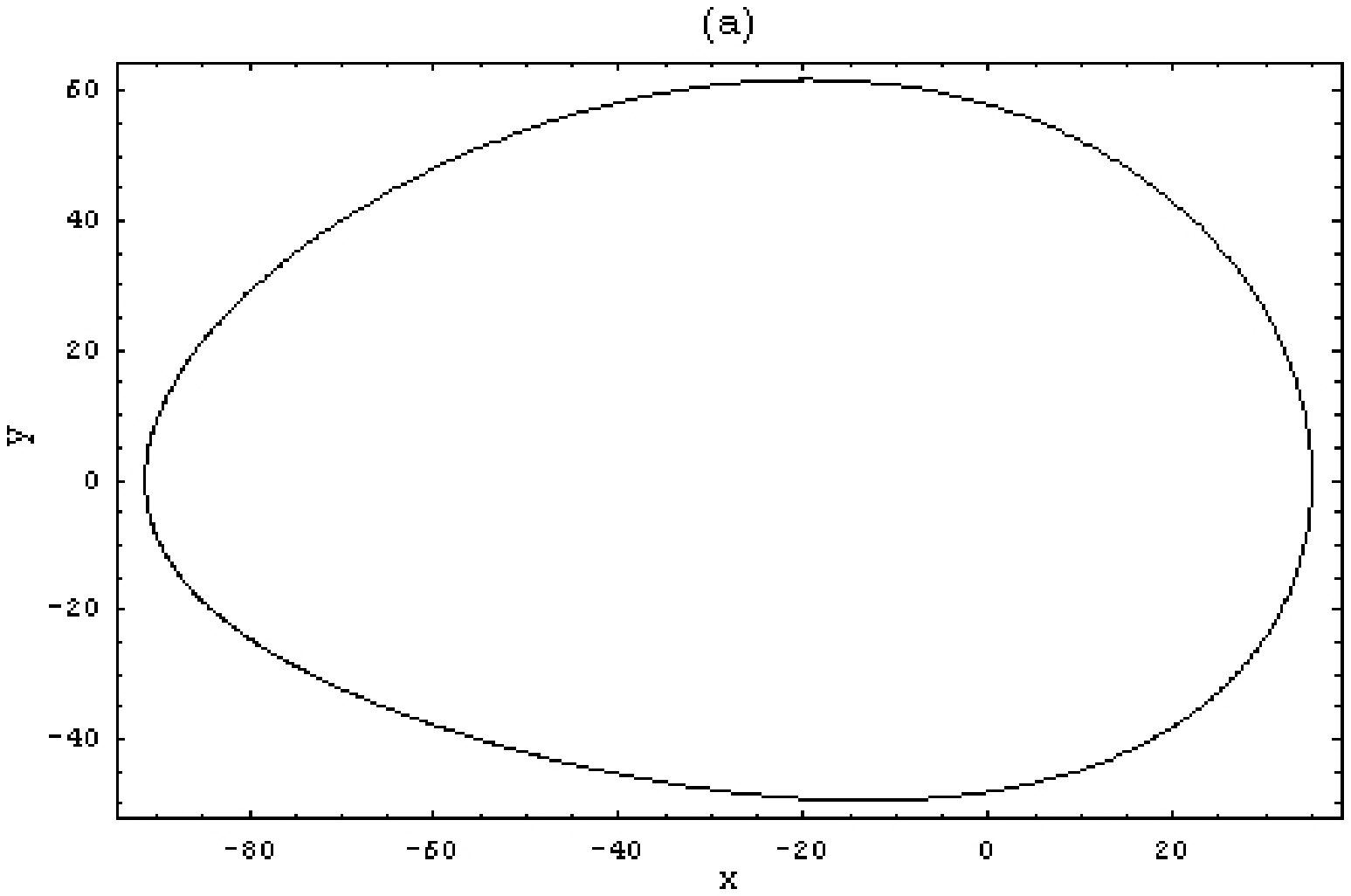,width=10.2cm,angle=0}
 \\
\vspace{1cm} ~
\\
\epsfig{file=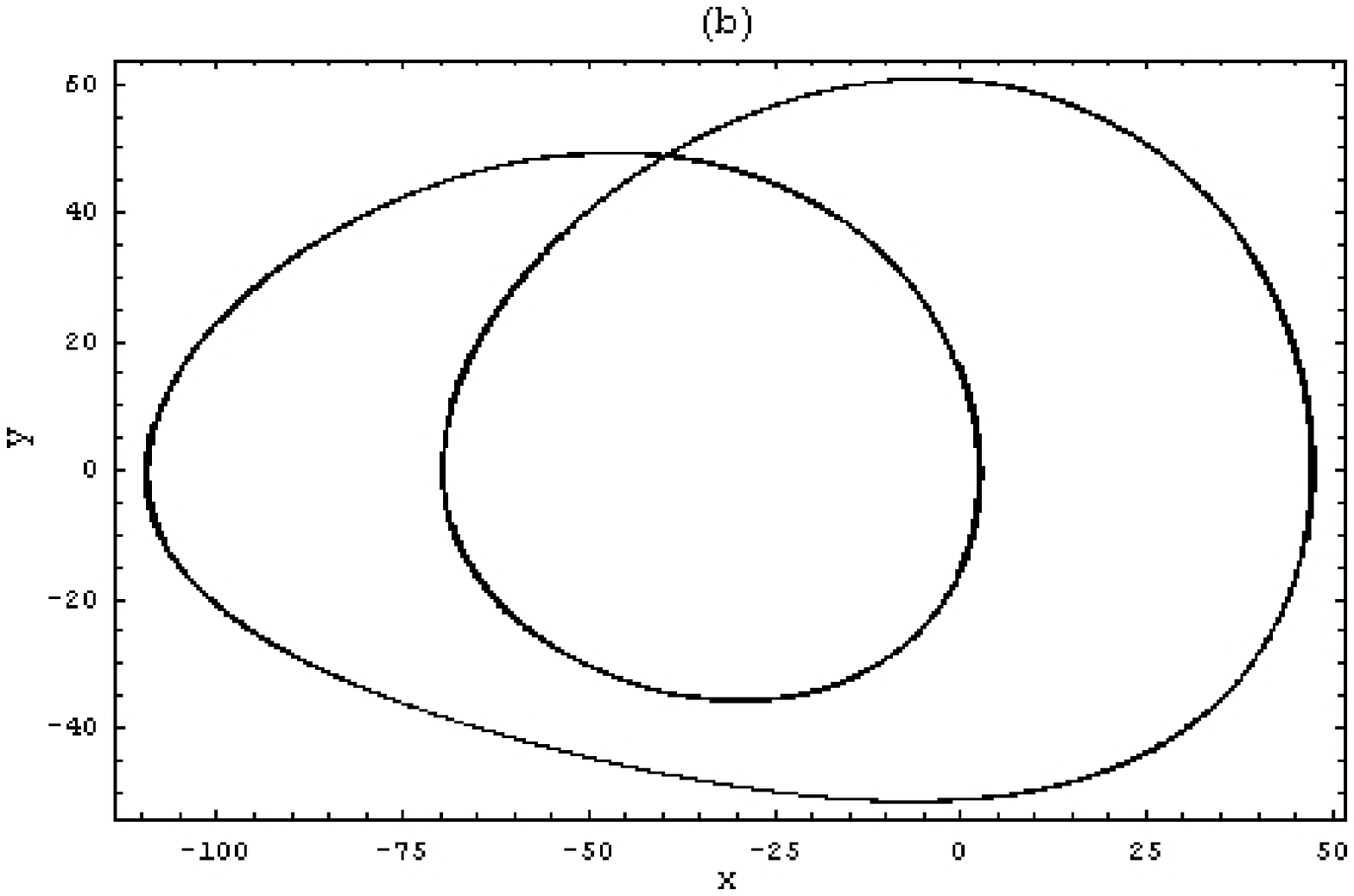,width=10.2cm,angle=0} 
\end{center} 

 \caption{ \label{fig4}
Stabilized period-1 (Fig \ref{fig4}a), period-2 (Fig \ref{fig4}b) and
period-4 (Fig \ref{fig4}c)
orbits of system (Eq. \ref{eq2})
using pulsive feedback technique with $\epsilon=-0.05$, -0.03 and -0.009,
respectively. A bifurcation diagram 
with pulsive feedback versus $\epsilon \in [-0.05,1]$.}

\end{figure}

\begin{figure}

\begin{center}
 \epsfig{file=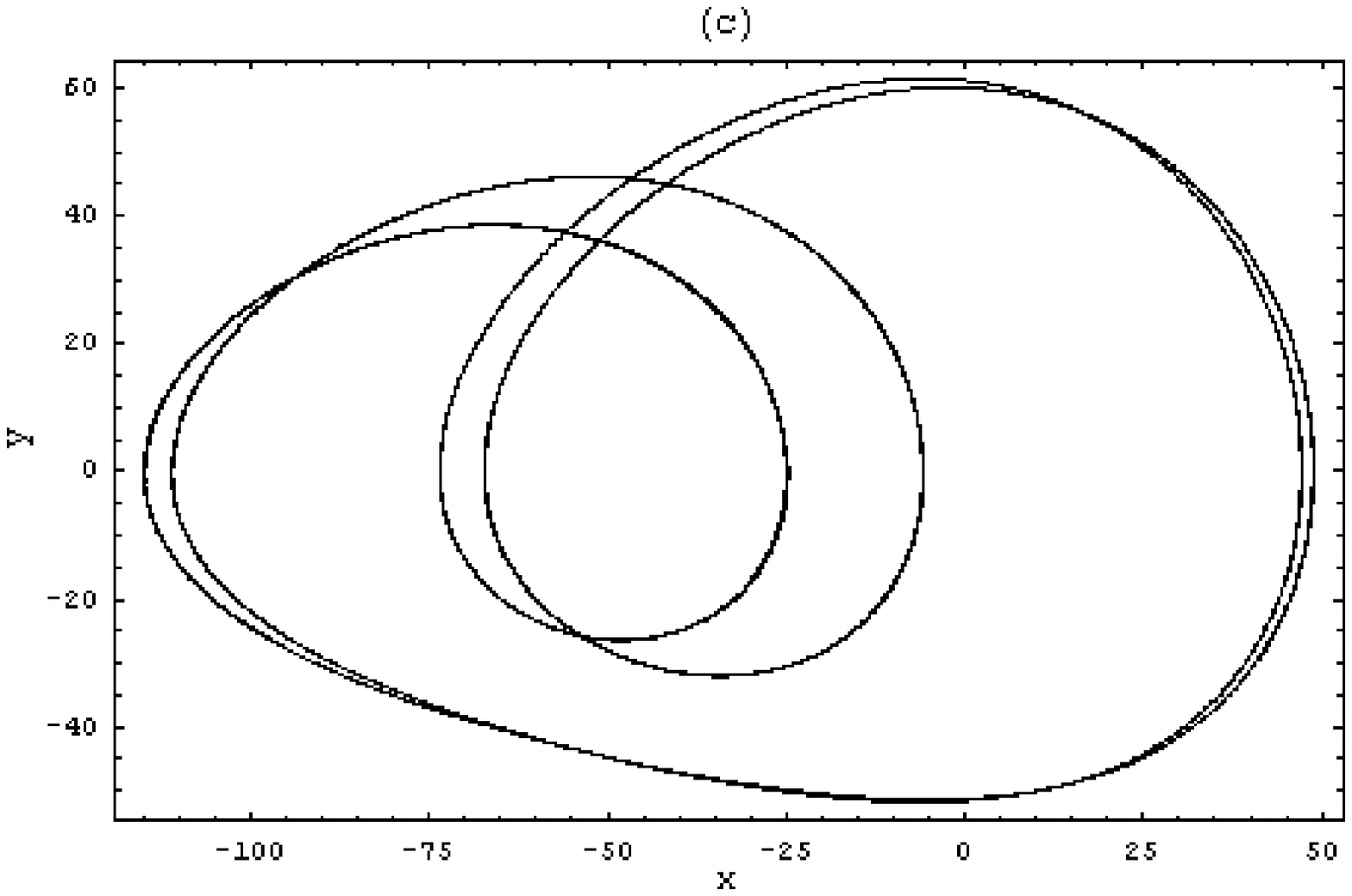,width=10.2cm,angle=0} \\
\vspace{1cm} ~
\\
\hspace{1.5cm}{\small (d)} \\
 \epsfig{file=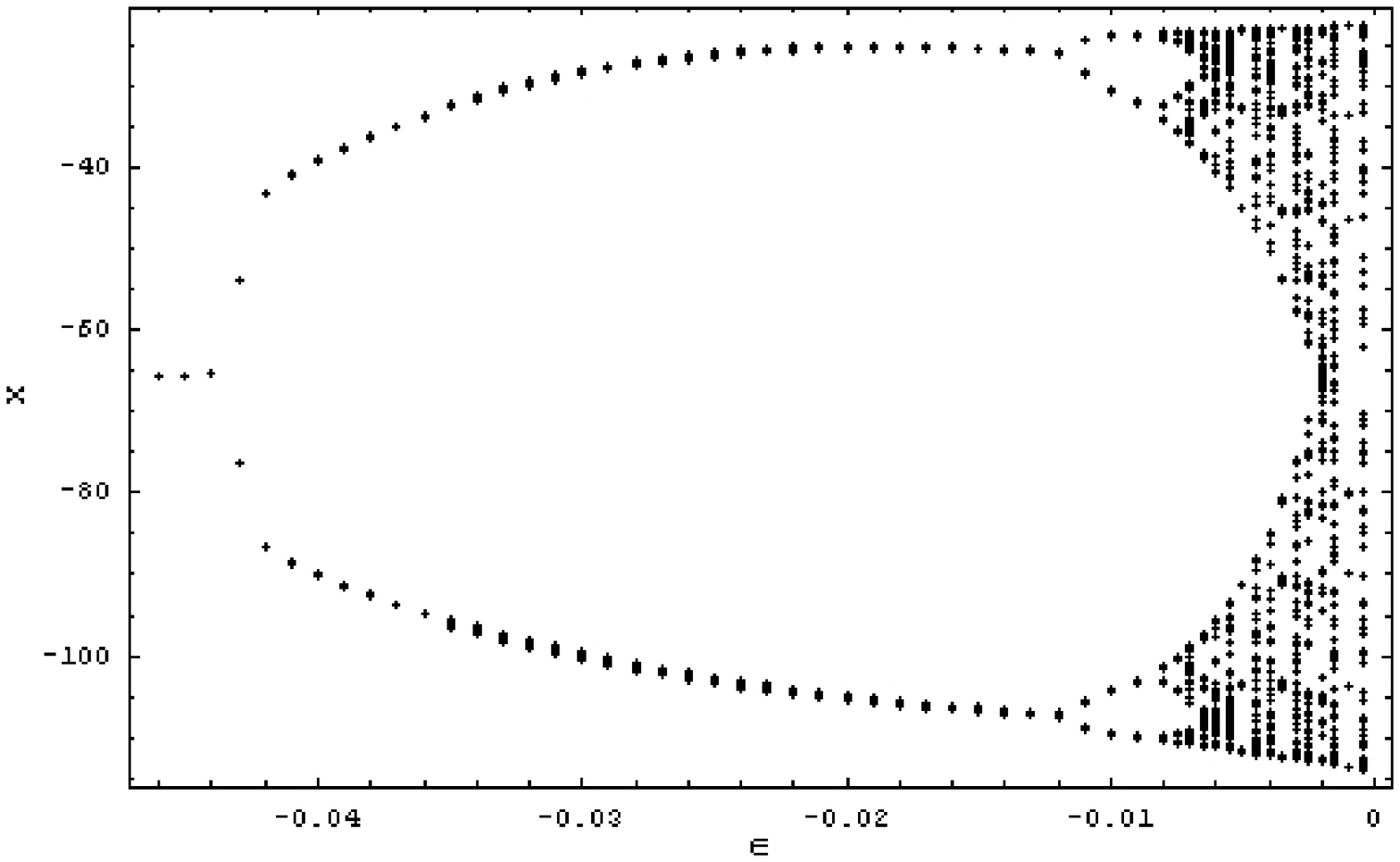,width=9.5cm,angle=0}
\end{center}

{Fig. 4 continuation.}
\end{figure}

\end{document}